\documentclass[sigconf]{nimeart}

\usepackage{xurl}

\usepackage{CJKutf8}
\usepackage{xcolor} 

\hypersetup{
  colorlinks=true,
  linkcolor=blue,
  urlcolor=blue,
  filecolor=blue,
  citecolor=blue
}

\newcommand{\orangei}{\textcolor{orange}{(i)}}
\newcommand{\orangeii}{\textcolor{orange}{(ii)}}
\newcommand{\orangeiii}{\textcolor{orange}{(iii)}}

\AtBeginDocument{%
  }

\setcopyright{cc}
\copyrightyear{2026}
\acmYear{2026}
\acmDOI{}

\acmConference[NIME '26]{International Conference on New Interfaces for Musical Expression}{June 23--26,
  2026}{London, UK}
\acmISBN{}

\begin{document}

\title{FXplorer: A Map-Based Interface for Exploratory Audio Effect Design}

\author{Annie Chu}
\email{anniechu@u.northwestern.edu}
\affiliation{%
  \institution{Northwestern University}
  \city{Chicago}
  \state{IL}
  \country{USA}
}

\author{Jason Brent Smith}
\email{jason.smith1@northwestern.edu}
\affiliation{%
  \institution{Northwestern University}
  \city{Chicago}
  \state{IL}
  \country{USA}
}

\author{Bryan Pardo}
\email{pardo@northwestern.edu}
\affiliation{%
  \institution{Northwestern University}
  \city{Chicago}
  \state{IL}
  \country{USA}
}

\renewcommand{\shortauthors}{Chu et al.}

\begin{abstract}
Audio effects (FX) shape sound in contemporary music practice. However, most interfaces present them as discrete modules and parameters that favor targeted adjustment over exploratory listening. This separation can make it difficult to build intuition about the broader space of possible transformations or to move fluidly between searching and refinement. We present \textit{FXplorer}, an interface that organizes audio effects within a perceptually informed 2D space, allowing sound transformations to be browsed as a continuous landscape rather than as isolated presets. By combining established spatial interaction approaches and interpretable DAW-style controls with recent embedding-based machine learning methods for similarity and semantic search, the system brings exploration and parameter refinement into a single workspace. \textit{FXplorer} supports composition, production, or performance by allowing users to edit and interpolate between effect presets interactively.
\end{abstract}

  
\begin{teaserfigure}
  \includegraphics[width=\textwidth]{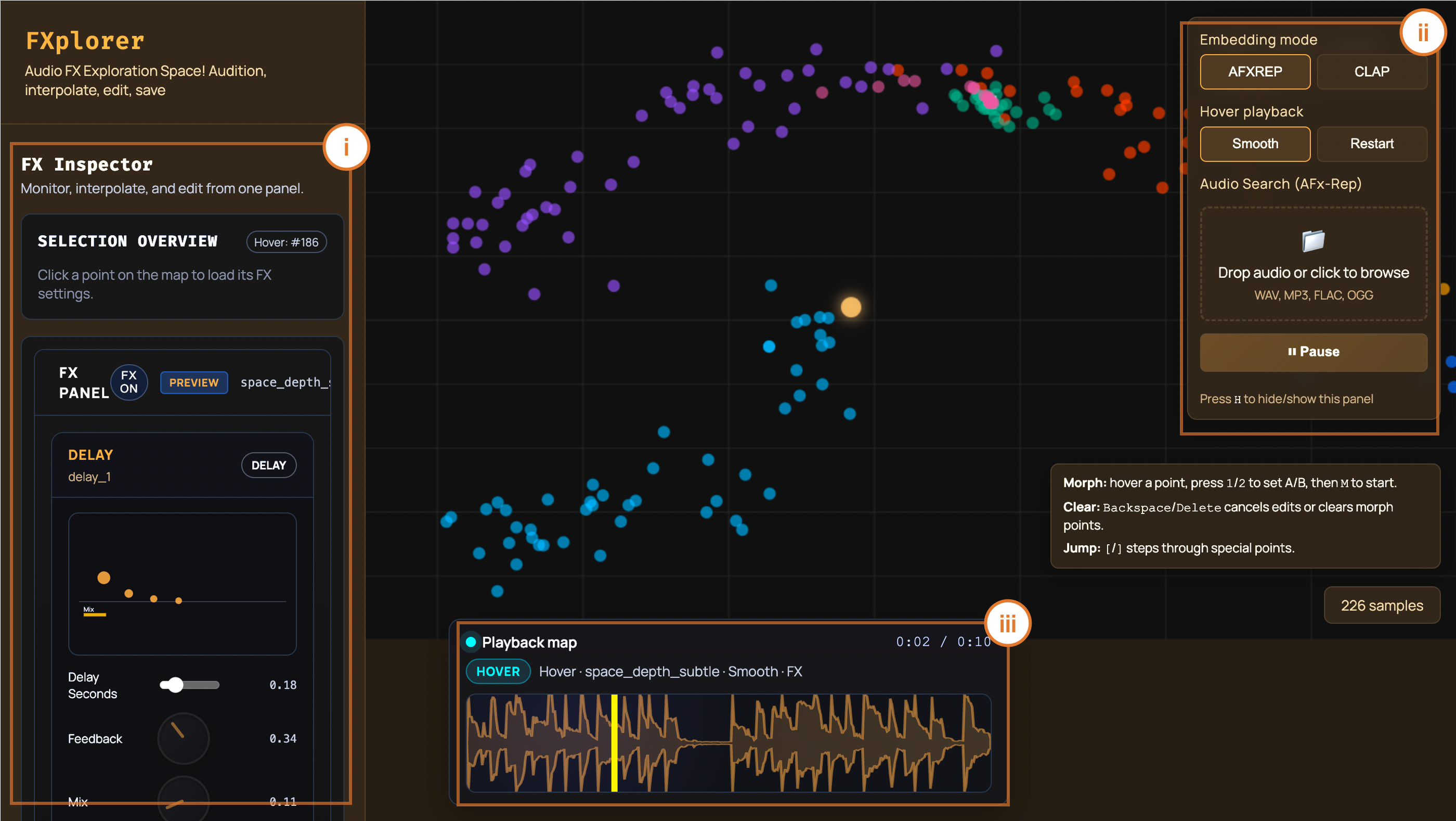}
  \caption{\textit{FXplorer}'s Explore Mode: Users primarily navigate via mouse (hover, click, pan, zoom) and keyboard shortcuts. Inspector panel \orangei{} for parameter editing, assigning interpolation endpoints, and viewing exact parameters for the currently selected effect variant. Floating control panel \orangeii{} for embedding mode switching (CLAP/AFx-Rep), hover playback settings, and semantic search inputs. Waveform monitor \orangeiii{} displays the currently playing audio with a moving playhead.}
  \Description{2D space of \textit{FXplorer}. Fx'd variations of an uploaded dry sound of created then embedded, allowing users to audition an `alternate universe` of input quickly and intuitively.}
  \label{fig:main}
\end{teaserfigure}

\maketitle

\section{Introduction}
Audio effects (FX) shape the timbral identity of recorded and performed sound, from subtle enhancement to radical transformation. The interfaces through which musicians access these possibilities often constrain creative exploration. For example, a producer seeking ``something warm and spacious'' must choose among discrete effect types (e.g., Reverb, EQ) and modify low-level parameters describing the underlying signal processing of a single effect (e.g., changing the ``Q'' of a filter) without clear guidance about what sonic possibilities exist and how they relate to one another. This reflects a broader gap between the expressive potential of audio FX and the limited exploratory affordances of current tools \cite{seetharaman2016audealize}.

Creative sound design involves both divergent exploration (discovering what is possible) and convergent refinement (achieving a specific sonic goal) \cite{tubb2014divergent}. While musicians frequently move fluidly between these modes, most digital audio workstations (DAWs) privilege refinement \cite{pardo2019learning}: users select effects from taxonomic menus and manipulate isolated parameters. The prior knowledge assumed by this workflow (e.g., which parameters matter) 
fails both newcomers building sonic intuitions and experienced practitioners in exploratory contexts.

Creators often begin with a vague intent \cite{delle2022semantic, Hertzmann2022TowardMCA} --perhaps a verbal description like ``something muddy'' or another audio sample that is ``muddy'' in their opinion-- and need a way to reach the space of ``muddy'' possibilities without necessarily knowing which FX chains or parameters to choose. After identifying a few promising transformations, they may explore the space between them, listening to how changes in mixing and effect parameters shape the sound. 

The key challenge in this workflow lies in the multidimensionality and nonlinearity of FX parameter spaces. Even simple effects expose vast combinatorial possibilities, and FX chains introduce order-dependent, perceptually-uneven interactions \cite{stasis2017audio}. Auditioning alternatives typically requires repeated cycles of parameter adjustment and rollback, discouraging rapid, low-commitment exploration \cite{delle2022semantic}. 
Preset libraries provide entry points but obscure the continuous sonic space between configurations \cite{seetharaman2016audealize}. Recent deep learning approaches enable semantic access to effects through text queries \cite{chu2025text2fx, doh2025can, ki2026fxsearcher} or audio examples \cite{steinmetz2024st}, but typically return isolated results rather than ongoing exploration.

Map-based interfaces have long been used in NIMEs to organize discrete sound corpora in perceptually meaningful 2D layouts \cite{fried2014audioquilt, font2017freesound, roma2019adaptive, garber2020audiostellar, dalri2025morphdrive}. Deep learning has further enabled synthesis-focused instruments such as RAVE \cite{caillon2021rave}, NSynth \cite{engel2017neural}, and related latent-space systems supporting continuous, perceptually organized exploration within similarly navigable 2D control spaces \cite{schwarz2006real, tan2017infinite, hantrakul2019klustr, engel2017neural, yang2019inspecting, xln2021xo, naradowsky2021amp}, including investigations of their gestural affordances \cite{Lunt2023Latent, zheng2025exploring}. These systems support low-commitment auditioning and reveal structure beyond categorical menus, yet typically operate on fixed-sound corpora or on abstract generative parameters. 

We present \textbf{\textit{FXplorer}}, a prototype interface for musical expression (NIME) for \textbf{exploratory audio FX design}. \textit{FXplorer} explores the question, ``what if audio effects could be navigated spatially, continuously and by ear, rather than through lists of modules and parameters?'' \textit{FXplorer} organizes hundreds of effect variants of a source sound into a perceptually organized two-dimensional space, where nearby points have timbral similarity. 
Musicians and performers can search this space using text or audio examples 
and 
explore how their changes reposition them within the sonic landscape. \textit{FXplorer} is designed to \textbf{support both divergent exploration and convergent refinement}, making FX parameter space navigable without sacrificing interpretability. 

\section{Design Requirements for Audio Effect Exploration}
Based on the challenges outlined above—navigating high-dimensional effect spaces, supporting rapid auditioning, connecting parameter changes to perceived sound, and enabling search through listening-based goals—we derive four design requirements for exploratory audio FX interfaces.


\textbf{DR1: Perceptually-organized exploration space.} Effect variants should be organized into a 2D layout where proximity reflects timbral similarity of transformations, revealing relationships beyond categorical FX hierarchies. 

\textbf{DR2: Semantic entry points into navigable space}. Text and audio queries should situate users within a continuous map rather than isolated retrieval results, enabling ``ballpark'' entry and subsequent exploration of adjacent sonic territories.

\textbf{DR3: Low-commitment auditioning of transformations.} The interface should enable rapid auditioning of dozens of configurations without explicit selection or commitment, facilitating divergent exploration while preserving parameter interpretability for convergent refinement. When a user finds two points they like, they should be able to ``slingshot,'' or navigate between them quickly and continuously.


\textbf{DR4: Connect real-time parameter edits to perceptual context}. The system should scale parameters perceptually (logarithmic for frequency, exponential for time constants, linear for gain/mix) and visualize how changes map to perceptual space.


\textit{FXplorer} integrates these requirements through a hybrid architecture: pre-rendered variants organized via audio embeddings (DR1), dual-embedding search supporting text \& audio queries (DR2), rapid in-situ auditioning (DR3), and module-aware editing with synchronized spatial/parametric/auditory feedback (DR4).

\section{System Design \& Implementation}
\textit{FXplorer} is a locally-hosted web interface prototype for composition or real-time musicking, designed with novices and low setup in mind. It displays audio effect variants on a fluid, interactive 2D canvas (see Fig. \ref{fig:main}). 

\textit{FXplorer} follows a hybrid architecture: computationally expensive operations such as variant generation 
are performed offline, while 
audio playback and interaction are handled in the browser for low-latency exploration. It's contained within a Svelte\footnote{\url{https://svelte.dev/}} web application that loads precomputed coordinates and audio at startup. Tone.js\footnote{\url{http://tonejs.github.io/}} enables real-time FX processing in the browser, reconstructing audio on-the-fly by applying parameter configurations to the dry source. A Flask API handles on-demand operations in the backend: projecting new points, embedding text queries, matching audio examples.

\begin{table}[]
    \centering
    \begin{tabular}{|c|c|}
        \hline
        \textbf{Keyboard Shortcut} & \textbf{Function} \\ \hline
        1 & Set Interpolation Endpoint A \\
        2 & Set Interpolation Endpoint B \\
        {[ and ]} & Cycle through Saved Points \\
        B & Toggle FX Bypass \\
        M & Enter Interpolation Mode \\
        $\leftarrow$ / $\rightarrow$ & Adjust Interpolation Blend \\
        H & Toggle Floating Controls \\
        \texttt{Ctrl/Cmd \& +/-} & Zoom In/Out \\
        \hline
    \end{tabular}
    \caption{\textit{FXplorer}'s Keyboard Shortcuts}
    \label{tab:shortcuts}
\end{table}

\subsection{Sample Generation} \label{sec:samplegen}
\textit{FXplorer}'s backend operates in three stages: generation, embedding, and reduction.

\textbf{Generation.} The system first randomly samples parameter configurations for user-selected effect modules and renders the variants using Pedalboard \footnote{https://github.com/spotify/pedalboard} (fast, bit-accurate DSP).

\textbf{Embedding.} Each variant is encoded with two complementary embedding models capturing different notions of similarity. 
AFx-Rep \cite{steinmetz2024st}, an audio-only model, embeds and organizes sounds by perceived timbral characteristics, while CLAP (audio-text) \cite{wu2023large} aligns sounds with language descriptors (e.g., ``warm,'' ``metallic''). 
Each embedding kind is projected separately into its own 2D layout. Users can toggle between these views, switching the notion of similarity while exploring the same set of variants. This enables neighborhood browsing and text- or audio-driven search within a consistent interface.

\textbf{Reduction.} These high-dimensional embeddings (~512) are projected to 2D via PCA \cite{pearson1901liii} (or UMAP \cite{mcinnes2018umap}) for visualization. Unlike methods like t-SNE \cite{maaten2008visualizing}, PCA/UMAP supports out-of-sample projection, enabling runtime mapping of new points (edits, searches) without refitting the entire space. 

\section{Interaction Design}
\begin{figure}[h!]
    \centering
    \includegraphics[width=0.85\linewidth]{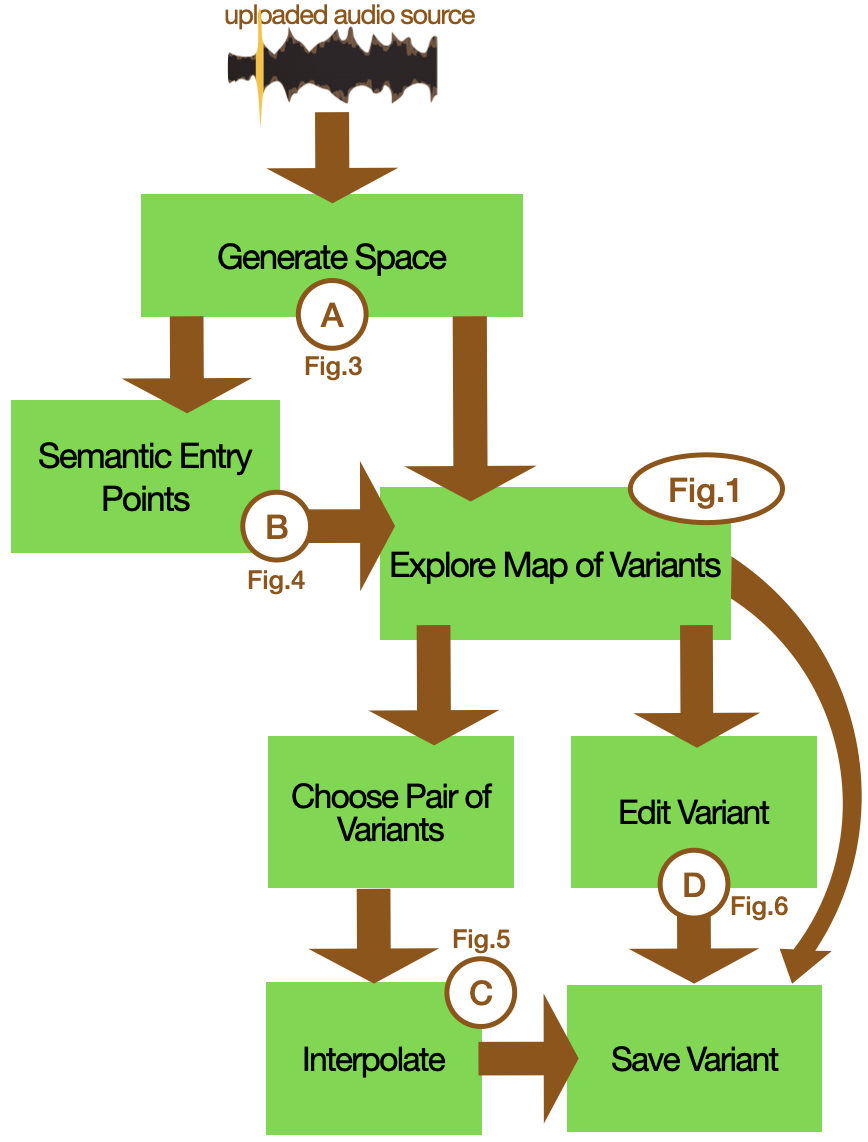}
    \caption{Overview of interaction workflows in \textit{FXplorer}. From an input audio source, the system constructs a semantic variant space that users navigate to explore alternatives, select and interpolate between variants, iteratively edit outcomes, and save results, supporting both exploratory (divergent) and refinement (convergent) processes.}    
    \label{fig:workflow}
\end{figure}

\textit{FXplorer} supports a creative workflow (Fig. \ref{fig:workflow}) where users generate a perceptually organized variant map and perform all browsing, search, interpolation, and refinement directly within this shared space before saving.

\subsection{Preparing the Exploration Space: Generating Variants (DR1)}
\begin{figure}[h!]
    \centering
    \includegraphics[width=0.9\linewidth]{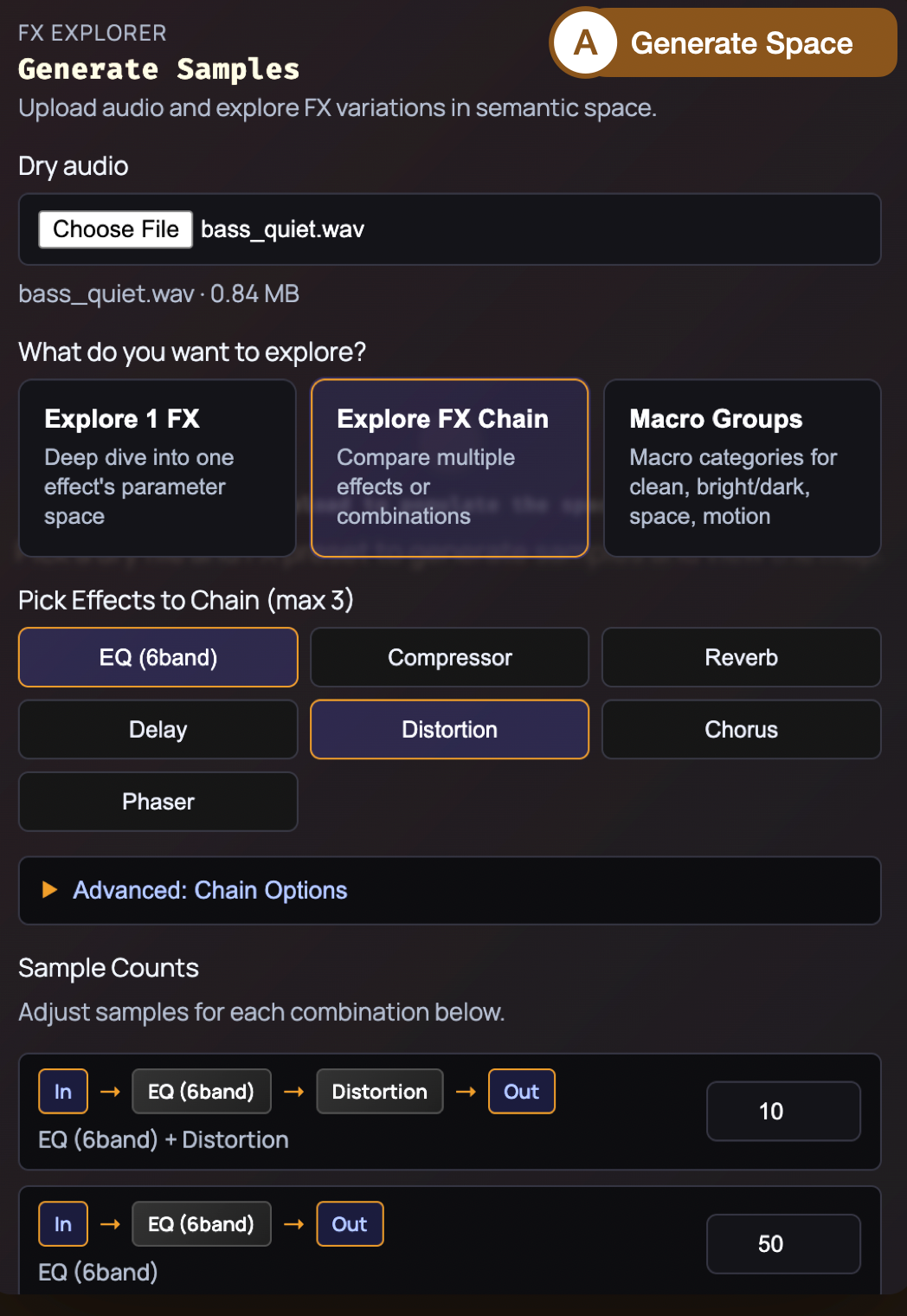}
    \caption{Sample generation in \textit{FXplorer}. Users upload dry audio, select an exploration mode (single, effect chain, or macro groups), choose effects to include and specify $n$ samples to generate for each configuration.}
    \label{fig:upload}
\end{figure}

Users may begin without a parameter plan, asking what transformations are possible for a source sound and generating a broad set of variations 
(Fig. \ref{fig:upload}).

Users begin by uploading dry audio, choose effect modules, and \textit{FXplorer} generates the variants that populate the map. \textit{FXplorer} was designed with short samples in mind (2-4s); Table \ref{tab:sample_length_benchmark} provides a benchmark on end-to-end wall-clock time. Available FX modules include 6-band parametric equalization (EQ), Compression, Reverb, Delay, Distortion, Chorus, and Phasing. Users can choose individual effects or combine up to three modules to explore interactions between effects.

\begin{table}[t]
\centering
\small
\begin{tabular}{lrrrrl}
\hline
Input Dur. & Generate & Embed & Reduce & Total & Device \\
\hline
2s  & 3.74 & 26.18 & 6.97 & 36.95 & CPU \\
2s  & 3.79 & 20.10 & 6.98 & 30.89 & GPU \\
\hline
4s  & 4.33 & 28.49 & 7.03 & 39.90 & CPU \\
4s  & 4.33 & 20.49 & 7.06 & 31.92 & GPU \\
\hline
10s & 6.10 & 35.54 & 7.25 & 49.00 & CPU \\
10s & 6.11 & 21.72 & 7.26 & 35.18 & GPU \\
\hline
\end{tabular}
\caption{Mean end-to-end wall-clock time (seconds) for the user-facing pipeline with a new dry input upload, \texttt{target\_samples}=100. Device affects only the \textit{Embed} stage; \textit{Embed} sums AFx-Rep and CLAP embedding time, \textit{Reduce} sums the two PCA reductions, and each row reports the mean over 3 runs.}
\label{tab:sample_length_benchmark}
\end{table}

\textit{FXplorer} samples parameter configurations for the selected modules and displays the resulting variants ($\approx100$, adjustable) as points on a 2D canvas. Each point represents a distinct effect configuration colored by FX chain type. The layout is organized by timbral similarity, so variants that sound alike cluster together regardless of the specific effects used. For example, a ``warm'' sound produced by low-pass filtering appears similar to a sound shaped with reverb, despite different underlying processes.

Once generated, this space serves as the canvas for subsequent browsing, searching, and editing.

\subsection{Entering the Space: Semantic Entry Points (DR2)}

\begin{figure}[h!]
    \centering
    \includegraphics[width= \linewidth]{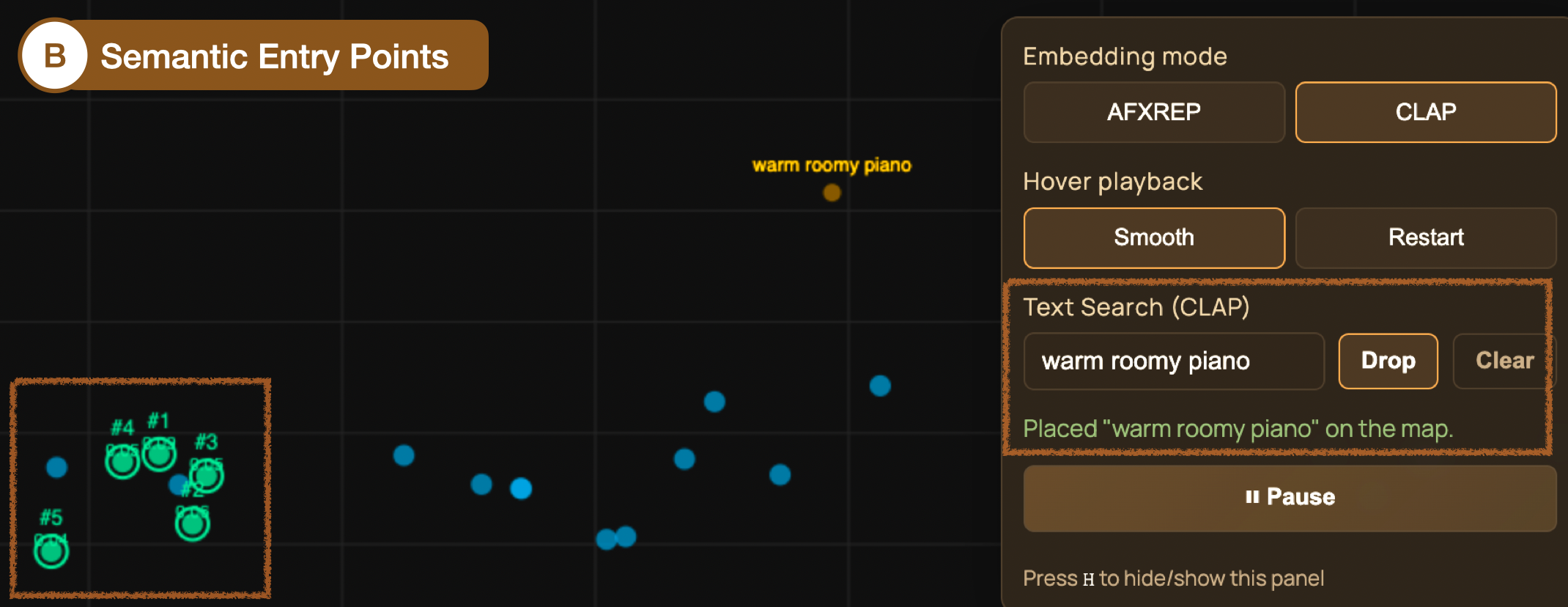}
    \caption{Semantic Entry Point Example (text): Results for ``warm roomy piano''. Embeds the actual query and highlights the top-5 matches.}
    \label{fig:semantic}
\end{figure}

Rather than scrolling through lists of isolated presets, users can enter the exploration space through semantic entry points. Starting from a rough sonic intention like ``warm and fuzzy,'' users can provide text queries or audio examples to search and highlight the most similar variants \textit{in-situ} within the existing map, situating results within their local neighborhood for immediate exploration (Fig. \ref{fig:semantic}). 

This search is computed by two complementary embeddings and cosine similarity: CLAP \cite{wu2023large} aligns sounds with descriptive language for text queries; AFx-Rep \cite{steinmetz2024st} measures timbral similarity for audio-to-audio matching. Users toggle between these semantic and timbral similarity modes while exploring the same variant set. 

\subsection{Exploring Variants by Rapid Audition \& Interpolation (DR3)}
Users explore the space by browsing and auditioning nearby variants. Moving the cursor over any point immediately plays the corresponding audio, with FX applied in real-time to the dry source. This enables rapid sampling of dozens of FX configurations. Clicking a point selects it for focused audition, begins audio looping, and exposes its parameters for editing in the Inspector.


\begin{figure}[h!]
    \centering
    \includegraphics[width=0.95\linewidth]{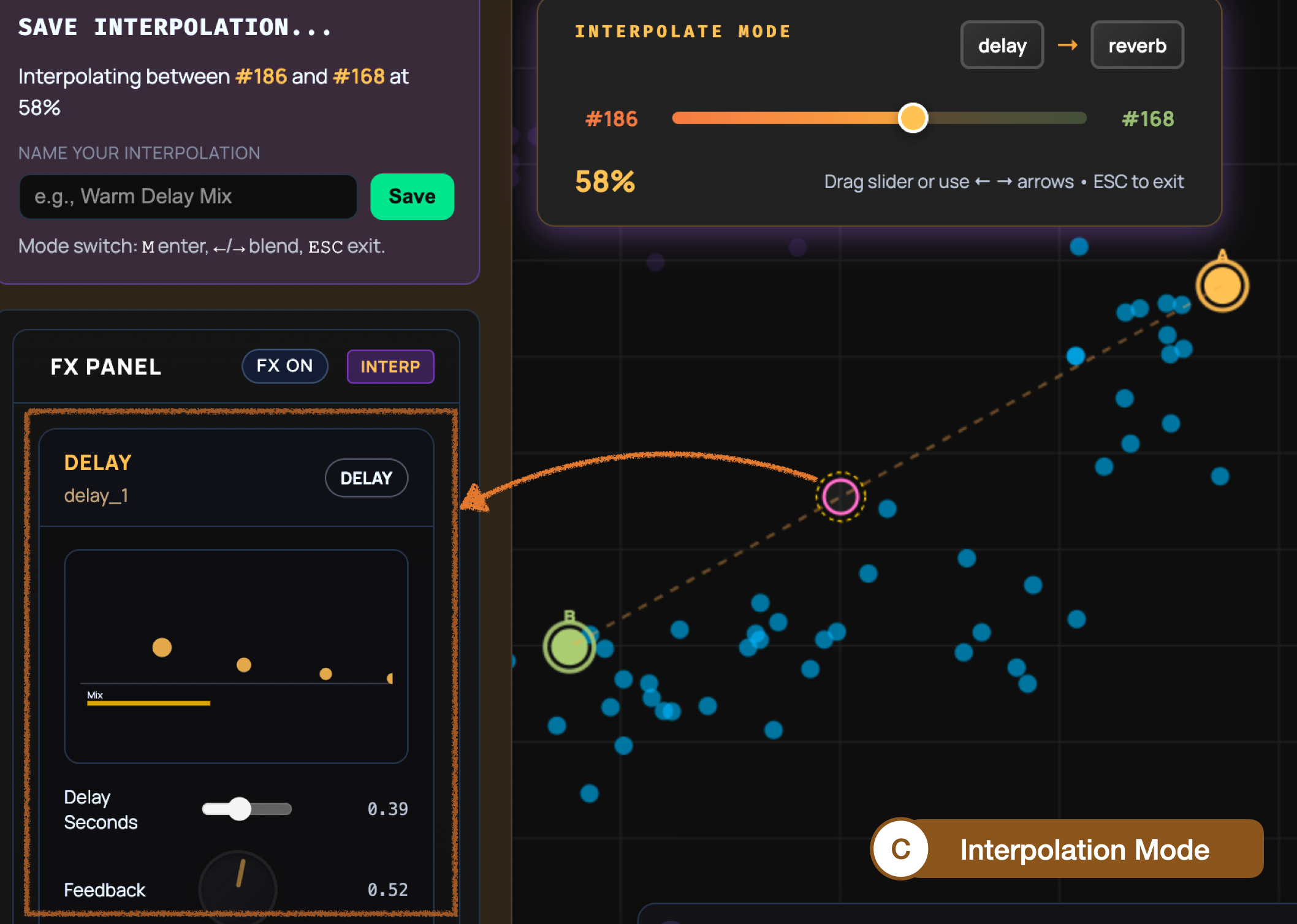}
    \caption{Interpolation Mode interface. Two samples from the exploration space are selected as points “A” and “B.” The user can quickly interpolate continuously between 
    endpoint parameters closest to point A (0\%) or B (100\%).}
    \label{fig:interpolation}
\end{figure}

When users identify two variants of interest, they can set them as endpoints via keyboard commands and continuously move between them to explore intermediate effects. Rather than crossfading audio, \textit{FXplorer} interpolates in parameter space: it blends the endpoint effect settings and applies the resulting configuration to the dry source in real time. When endpoints share the same effect chain, their parameters blend directly; different chains require manual endpoint reselection. The view zooms to focus on the endpoints, and the arrow keys slide across the interpolation line between the two endpoints (see Fig. \ref{fig:interpolation}). As users navigate, audio updates immediately; parameter values and their FX visuals (e.g., EQ curve, delay lines) adjust in the inspector panel; a ghost marker moves along the trajectory in 2D space, providing synchronized feedback. To support perceptually smooth transitions, parameters are scaled in effect-specific ways (e.g., logarithmic for frequency, exponential for time constants, linear for wet/dry). 

\subsection{Refining Ideas: Real-time Parameter Editing Feedback (DR4)}
\begin{figure}[h!]
    \centering
    \includegraphics[width=1.05\linewidth]{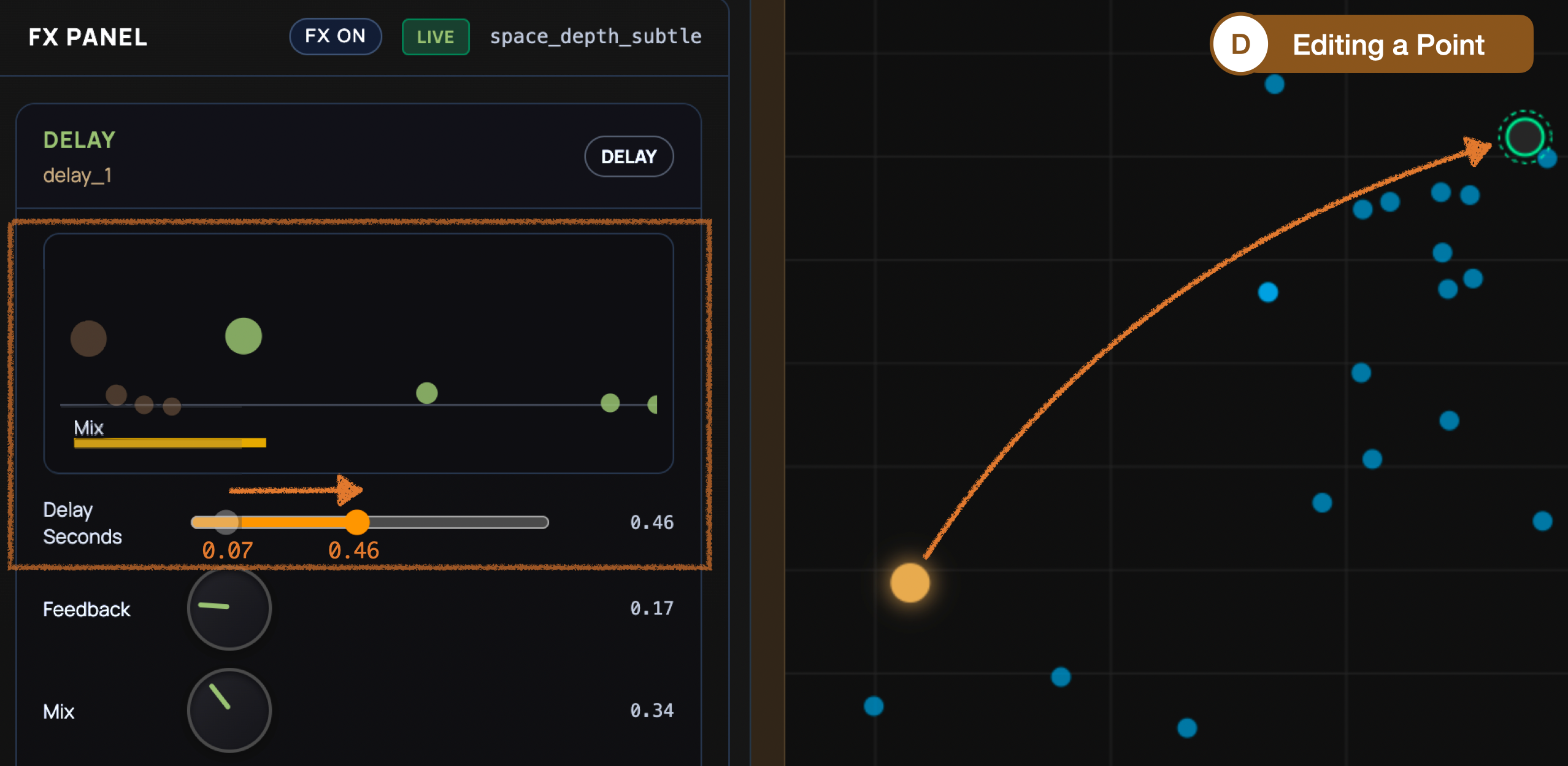}
    \caption{Variant editing selected point in the map (right) is opened in the FX panel (left), where users directly manipulate effect parameters (e.g., delay time, feedback, mix) to iteratively refine the sound and save an updated variant.}    
    \label{fig:edit}
\end{figure}

After finding a promising region or variant, users can refine a sound directly by adjusting its effect parameters in the Inspector. Edits are applied to the dry source in real time, allowing immediate auditory feedback while the sound loops. To maintain context within the exploration space, \textit{FXplorer} projects edited configurations back into the 2D layout, enabled by re-embedding and PCA's out-of-sample projection. A ``ghost'' point marker shows where modified settings would lie on the map, linking parameter changes to their spatial position. Large movements indicate substantial timbral change, while small shifts reflect subtle adjustments (Fig \ref{fig:edit}). This feedback aims to help users develop intuition about how parameter edits relate to sonic transformations.

\section{Conclusions}
We present \textit{FXplorer,} a prototype interface that reframes audio effect design from a sequence of discrete parameter adjustments to navigation through a continuous landscape of transformations. By unifying rapid auditioning, interpolation, semantic search, and direct parameter editing within a single audio-visual space, the system enables fluid movement between exploration and refinement.

This work extends the NIME tradition of map-based navigational interfaces to audio FX. Unlike systems that operate over fixed corpora or abstract 
parameters, \textit{FXplorer} maintains explicit, DAW-compatible controls, aiming to balance exploratory interaction 
with practical production needs. 
Future work could evaluate how these spatial and semantic affordances influence sound design strategies, 
exploring larger or adaptive variant sets. 

\section{Ethical Standards}
The system uses machine learning only for similarity estimation and retrieval through discriminative embeddings; it does not generate audio or train models during use. All sounds derive from user-provided material and transparent effect parameters. Processing occurs locally; embedding and projection introduce some computational overhead, though the footprint is substantially lower than generative approaches. We acknowledge the potential for some artists to feel embedding-based similarity suggestions stifle their creativity. Our goal is to assist artists in this exploratory process; we believe designing \textit{FXplorer} with creators in mind overcomes potential adverse impact on human creativity. 

\begin{acks}
This work was supported by NSF Award Number 2300633.
\end{acks}

\bibliographystyle{ACM-Reference-Format}
\bibliography{references}

@inproceedings{tubb2014divergent,
 address = {London, United Kingdom},
 author = {Robert Tubb and Simon Dixon},
 booktitle = {Proceedings of the International Conference on New Interfaces for Musical Expression},
 doi = {10.5281/zenodo.1178967},
 issn = {2220-4806},
 month = {June},
 pages = {227--232},
 publisher = {Goldsmiths, University of London},
 title = {The Divergent Interface: Supporting Creative Exploration of Parameter Spaces},
 url = {http://www.nime.org/proceedings/2014/nime2014_415.pdf},
 year = {2014}
}

@inproceedings{stasis2017audio,
  title={Audio processing chain recommendation},
  author={Stasis, Spyridon and Jillings, Nicholas and Enderby, Sean and Stables, Ryan},
  booktitle={Proceedings of the 20th International Conference on Digital Audio Effects,(Edinburgh, UK)},
  year={2017}
}

@article{delle2022semantic,
  title={Semantic models of sound-driven design: Designing with listening in mind},
  author={Delle Monache, Stefano and Misdariis, Nicolas and {\"O}zcan, Elif},
  journal={Design Studies},
  volume={83},
  pages={101134},
  year={2022},
  publisher={Elsevier}
}

@inproceedings{chu2025text2fx,
  title={Text2fx: Harnessing clap embeddings for text-guided audio effects},
  author={Chu, Annie and O’Reilly, Patrick and Barnett, Julia and Pardo, Bryan},
  booktitle={ICASSP 2025-2025 IEEE International Conference on Acoustics, Speech and Signal Processing (ICASSP)},
  pages={1--5},
  year={2025},
  organization={IEEE}
}

@article{doh2025can,
  title={Can Large Language Models Predict Audio Effects Parameters from Natural Language?},
  author={Doh, Seungheon and Koo, Junghyun and Mart{\'\i}nez-Ram{\'\i}rez, Marco A and Liao, Wei-Hsiang and Nam, Juhan and Mitsufuji, Yuki},
  journal={arXiv preprint arXiv:2505.20770},
  year={2025}
}

@inproceedings{ki2026fxsearcher,
  title={FxSearcher: gradient-free text-driven audio transformation},
  author={Ki, Hojoon and Kim, Jongsuk and Kwon, Minchan and Kim, Junmo},
  booktitle={ICASSP 2026-2026 IEEE International Conference on Acoustics, Speech and Signal Processing (ICASSP)},
  pages={15462--15466},
  year={2026},
  organization={IEEE}
}

@article{steinmetz2024st,
  title={St-ito: Controlling audio effects for style transfer with inference-time optimization},
  author={Steinmetz, Christian J and Singh, Shubhr and Comunit{\`a}, Marco and Ibnyahya, Ilias and Yuan, Shanxin and Benetos, Emmanouil and Reiss, Joshua D},
  journal={arXiv preprint arXiv:2410.21233},
  year={2024}
}

@article{seetharaman2016audealize,
  title={Audealize: Crowdsourced audio production tools},
  author={Seetharaman, Prem and Pardo, Bryan},
  journal={Journal of the Audio Engineering Society},
  volume={64},
  number={9},
  pages={683--695},
  year={2016},
  publisher={Audio Engineering Society}
}

@inproceedings{font2017freesound,
  title={Freesound explorer: make music while discovering freesound!},
  author={Font, Frederic and Bandiera, Giuseppe},
  booktitle={Proceedings of the 3rd Web Audio Conference},
  year={2017}
}

@inproceedings{fried2014audioquilt,
 address = {London, United Kingdom},
 author = {Ohad Fried and Zeyu Jin and Reid Oda and Adam Finkelstein},
 booktitle = {Proceedings of the International Conference on New Interfaces for Musical Expression},
 doi = {10.5281/zenodo.1178766},
 issn = {2220-4806},
 month = {June},
 pages = {281--286},
 publisher = {Goldsmiths, University of London},
 title = {AudioQuilt: {2D} Arrangements of Audio Samples using Metric Learning and Kernelized Sorting},
 url = {http://www.nime.org/proceedings/2014/nime2014_315.pdf},
 year = {2014}
}

@inproceedings{roma2019adaptive,
 address = {Porto Alegre, Brazil},
 author = {Gerard Roma and Owen Green and Pierre Alexandre Tremblay},
 booktitle = {Proceedings of the International Conference on New Interfaces for Musical Expression},
 doi = {10.5281/zenodo.3672976},
 editor = {Marcelo Queiroz and Anna Xambó Sedó},
 issn = {2220-4806},
 month = {June},
 pages = {313--318},
 publisher = {UFRGS},
 title = {Adaptive Mapping of Sound Collections for Data-driven Musical Interfaces},
 url = {http://www.nime.org/proceedings/2019/nime2019_paper060.pdf},
 year = {2019}
}

@inproceedings{pardo2019learning,
  title={Learning to build natural audio production interfaces},
  author={Pardo, Bryan and Cartwright, Mark and Seetharaman, Prem and Kim, Bongjun},
  booktitle={Arts},
  volume={8},
  number={3},
  pages={110},
  year={2019},
  organization={MDPI}
}

@inproceedings{engel2017neural,
  title={Neural audio synthesis of musical notes with wavenet autoencoders},
  author={Engel, Jesse and Resnick, Cinjon and Roberts, Adam and Dieleman, Sander and Norouzi, Mohammad and Eck, Douglas and Simonyan, Karen},
  booktitle={International conference on machine learning},
  pages={1068--1077},
  year={2017},
  organization={PMLR}
}

@inproceedings{Hertzmann2022TowardMCA,
  title={Toward Modeling Creative Processes for Algorithmic Painting},
  author={Aaron Hertzmann},
  booktitle={ICCC},
  year={2022},
  url={https://api.semanticscholar.org/CorpusId:248506151}
}

@article{yang2019inspecting,
  title={Inspecting and interacting with meaningful music representations using VAE},
  author={Yang, Ruihan and Chen, Tianyao and Zhang, Yiyi and Xia, Gus},
  journal={arXiv preprint arXiv:1904.08842},
  year={2019}
}

@article{Lunt2023Latent,
	author = {Lunt, Alexander and Trump, Sebastian},
	journal = {AIMC 2023},
	year = {2023},
	month = {aug 29},
	note = {https://aimc2023.pubpub.org/pub/zgc5j7ha},
	publisher = {},
	title = {Latent {Space} {Explorer}},
}

@article{zheng2025exploring,
  title={Exploring gestural affordances in audio latent space navigation},
  author={Zheng, Shuoyang Jasper and Xamb{\'o} Sed{\'o}, Anna and Bryan-Kinns, Nick},
  journal={Frontiers in Computer Science},
  volume={7},
  pages={1575202},
  year={2025},
  publisher={Frontiers Media SA}
}

@inproceedings{wu2023large,
  title={Large-scale contrastive language-audio pretraining with feature fusion and keyword-to-caption augmentation},
  author={Wu, Yusong and Chen, Ke and Zhang, Tianyu and Hui, Yuchen and Berg-Kirkpatrick, Taylor and Dubnov, Shlomo},
  booktitle={ICASSP 2023-2023 IEEE International Conference on Acoustics, Speech and Signal Processing (ICASSP)},
  pages={1--5},
  year={2023},
  organization={IEEE}
}

@article{caillon2021rave,
  title={RAVE: A variational autoencoder for fast and high-quality neural audio synthesis},
  author={Caillon, Antoine and Esling, Philippe},
  journal={arXiv preprint arXiv:2111.05011},
  year={2021}
}

@article{pearson1901liii,
  title={LIII. On lines and planes of closest fit to systems of points in space},
  author={Pearson, Karl},
  journal={The London, Edinburgh, and Dublin philosophical magazine and journal of science},
  volume={2},
  number={11},
  pages={559--572},
  year={1901},
  publisher={Taylor \& Francis}
}

@article{mcinnes2018umap,
  title={Umap: Uniform manifold approximation and projection for dimension reduction},
  author={McInnes, Leland and Healy, John and Melville, James},
  journal={arXiv preprint arXiv:1802.03426},
  year={2018}
}

@article{maaten2008visualizing,
  title={Visualizing data using t-SNE},
  author={Maaten, Laurens van der and Hinton, Geoffrey},
  journal={Journal of Machine Learning Research},
  volume={9},
  number={Nov},
  pages={2579--2605},
  year={2008}
}

@inproceedings{dalri2025morphdrive,
    author = "Dal Rí, Francesco Ardan and Stefani, Domenico and Turchet, Luca and Conci, Nicola",
    title = "{MorphDrive: Latent Conditioning for Cross-Circuit Effect Modeling and a Parametric Audio Dataset of Analog Overdrive Pedals}",
    booktitle = "Proceedings of the 28-th Int. Conf. on Digital Audio Effects (DAFx25)",
    editor = "Gabrielli, L. and Cecchi, S.",
    location = "Ancona, Italy",
    eventdate = "2025-09-02/2025-09-05",
    year = "2025",
    month = "Sept",
    publisher = "",
    issn = "2413-6689",
    doi = "",
    pages = ""
}

@inproceedings{naradowsky2021amp,
  title={Amp-space: A large-scale dataset for fine-grained timbre transformation},
  author={Naradowsky, Jason},
  booktitle={2021 24th International Conference on Digital Audio Effects (DAFx)},
  pages={57--64},
  year={2021},
  organization={IEEE}
}

@inproceedings{garber2020audiostellar,
  title={AudioStellar, an open source corpus-based musical instrument for latent sound structure discovery and sonic experimentation},
  author={Garber, Leandro and Ciccola, Tom{\'a}s and Amusategui, Juan Cruz},
  booktitle={Proceedings of ICMC},
  year={2020}
}

@inproceedings{schwarz2006real,
  title={Real-time corpus-based concatenative synthesis with catart},
  author={Schwarz, Diemo and Beller, Gr{\'e}gory and Verbrugghe, Bruno and Britton, Sam},
  booktitle={9th International Conference on Digital Audio Effects (DAFx)},
  pages={279--282},
  year={2006}
}

@misc{tan2017infinite,
  title={Infinite Drum Machine },
  author={Tan, Manny and McDonald, Kyle},
  year={2017},
  howpublished = {[Online] \url{https://experiments.withgoogle.com/ai/drum-machine/}}
}

@misc{hantrakul2019klustr,
    title={lamtharnhantrakul/klustr},
    author={Hantrakul, L. H.},
    year={2017},
    howpublished = {[Online]},
    url = {https://github.com/lamtharnhantrakul/klustr}
}

@misc{xln2021xo,
    author = {{XLN Audio}},
    title = {{XO - XLN Audio}},
    year = {2021},
    howpublished = {[Online]},
    url = {https://www.xlnaudio.com/products/xo}
}

\end{document}